\documentclass[conference]{IEEEtran}
\IEEEoverridecommandlockouts
\usepackage{cite}
\usepackage{amsmath,amssymb,amsfonts}
\usepackage{algorithmic}
\usepackage{graphicx}
\usepackage{textcomp}
\usepackage{hyperref}
\usepackage{xcolor}
\usepackage{amsmath}
\def\BibTeX{{\rm B\kern-.05em{\sc i\kern-.025em b}\kern-.08em
    T\kern-.1667em\lower.7ex\hbox{E}\kern-.125emX}}
\begin{document}

\title{memorAIs: an Optical Character Recognition and Rule-Based Medication Intake Reminder-Generating Solution\\

}

\author{\IEEEauthorblockN{Eden Shaveet}
\IEEEauthorblockA{\textit{Dept. Computer Science} \\
\textit{Columbia University}\\
New York, USA \\
ems2349@columbia.edu}
\and
\IEEEauthorblockN{Utkarsh Singh}
\IEEEauthorblockA{\textit{Dept. Computer Science} \\
\textit{Stony Brook University}\\
Stony Brook, USA \\
utkarsh.singh@stonybrook.edu}
\and
\IEEEauthorblockN{Nicholas Assaderaghi}
\IEEEauthorblockA{\textit{Dept. Computer Science} \\
\textit{Columbia University}\\
New York, USA \\
nra2130@columbia.edu}
\and
\IEEEauthorblockN{Maximo Librandi}
\IEEEauthorblockA{\textit{Dept. Computer Science} \\
\textit{Columbia University}\\
New York, USA \\
ml5014@columbia.edu}
}

\maketitle

\begin{abstract}
Memory-based medication non-adherence is an unsolved problem that is responsible for considerable disease burden in the United States. Digital medication intake reminder solutions with minimal onboarding requirements that are usable at the point of medication acquisition may help to alleviate this problem by offering a low barrier way to help people remember to take their medications. In this paper, we propose memorAIs, a digital medication intake reminder solution that mitigates onboarding friction by leveraging optical character recognition strategies for text extraction from medication bottles and rule based expressions for text processing to create configured medication reminders as local device calendar invitations. We describe our ideation and development process, as well as limitations of the current implementation. memorAIs was the winner of the Patient Safety award at the 2023 Columbia University DivHacks Hackathon, presented by the Patient Safety Technology Challenge, sponsored by the Pittsburgh Regional Health Initiative.
\end{abstract}

\section{Introduction}
\subsection{Medication Adherence}
Medication adherence is the extent to which a patient follows a stipulatory medication treatment plan \cite{b1}. Over half of U.S. adults do not take medications as prescribed, which is responsible for an estimated 33\%–69\% of hospital admissions and 125,000 deaths annually \cite{b1}. When medications are appropriately prescribed and made accessible to patients in a timely fashion, the value proposition of outpatient medication adherence is seen across several disease progression use cases \cite{b2}, \cite{b3}. For instance, adherence to disease-modifying medication regimens and therapies for the treatment and control of Multiple Sclerosis in early stages of disease onset is associated with decreased disease severity over time \cite{b4}. 

As more medications are taken outside of healthcare settings than in U.S. hospitals and clinics combined, better home medication management and intake practices may be critical to improving medication adherence overall \cite{b5}. In 2003, the World Health Organization (WHO) corroborated this claim by stating that increased effectiveness of medication adherence interventions may have “a far greater impact on the health of the population than any improvement in specific medical treatments” \cite{b6}.

\subsection{Barriers to Adherence}
There are several well-documented barriers to medication adherence, including high costs of medications and routine care \cite{b7}; lack of access to prescribers and pharmacies \cite{b8}; medical mistrust of clinical guidance or medications \cite{b9}; and memory-based adherence problems \cite{b10}. Memory-based non-adherence is a prevalent reason for home medication non-adherence, specifically for patients with cognitive impairments \cite{b11}, and has led to the development of several reminder-based solutions, including user-facing digital solutions \cite{b12}. However, of the several digital reminder-based solutions on the market, none have demonstrated significant improvements in adherence among their users \cite{b13}\cite{b14}\cite{b15}. While there may be several reasons for this lack of efficacy, one proposed reason is under-utilization.

Known barriers to digital health solution utilization at large include low digital literacy among users \cite{b16}, meaning that users with limited experience with other digital tools may find digital health tools equally difficult to use, and onboarding friction \cite{b17}, meaning that the effort exerted to configure the solution outweighs the perceived benefits of its use. For instance, patients with complex medication regimens, who would be required to manually input information for each medication, along with their corresponding routine details, before being able to meaningfully use a digital medication reminder application, may perceive the setup of the application as cumbersome, potentially discouraging its use.

In an effort to address the challenge of onboarding friction, we developed a solution that offers the benefit of reminder-based solutions while mitigating onboarding friction attributed to user-facing configuration tasks. In this paper, we present a prototype implementation of an automated medication reminder application, memorAIs, that extracts text from prescription medication containers using optical character recognition models and identifies medication intake time, duration, and frequency information using text classification-based strategies. The application returns this information in the form of an ics (iCalendar) file directly to a user’s device, eliminating the need for manual configuration.

memorAIs was conceptualized, designed, developed, and implemented in less than 36 hours at the 2023 DivHacks Hackathon at Columbia University. It was awarded the first-place Patient Safety Award from the Patient Safety Technology Challenge, sponsored by the Pittsburgh Regional Health Initiative.

\subsection{Solution Overview}
Our application prototype, memorAIs, is a platform that allows users to scan their on-the-bottle prescription intake directions using their device’s camera to generate and download a ics (iCalendar) file pre-configured with intake frequency, duration, and time information straight to their device in use. ics files are calendar files stored in a universal calendar format used by all major email and calendar programs, including Microsoft Outlook, Google Calendar, and Apple Calendar \cite{b18}.

To use the application, a user can access our front-end interface and has the option to either upload a photo of their prescription label, including the intake directions from the pharmacy, or capture one using their device's camera. Our system processes the photo using the following methods:

\begin{enumerate}
  \item Optical character recognition (OCR) models, which extract text from .png and .jpg files and return relevant substrings given text probability and character coordinates.
  \item Regular-expression-based text interpretation that associates keywords with the specific frequency, duration, and time details required to create a calendar reminder.
\end{enumerate}

The extracted details are then used to generate and export an ics file directly to the user's current device. All they need to do is download the file, and their medication reminders will be seamlessly integrated into a calendar of their choice.

\section{Methodology}
\subsection{Solution Design}
We began by considering patient-facing medication-acquisition and intake process flows and the point at which a digital reminder may be useful. For example, some patients may acquire their medications in-person from a pharmacy. Others may use a mail delivery service. Others may rely on a third-party service for hand-delivery. To minimize opportunities for memory-based non-adherence, we determined that our solution needed to be usable at the point of medication acquisition in any of these circumstances. In order for the solution to be usable at the point of medication acquisition, the solution’s requirements of a user would need to be minimal, quick, and easy.

Recognizing that prescription medication bottles typically provide intake instructions and that individuals using digital intake reminder solutions often possess devices such as smartphones equipped with cameras, we opted to create a solution integrating these elements. The optimal process flow we envisioned for the solution depended on patients capturing an image of the medication intake instructions and developing an underlying program that would extract, process, and return the information contained therein as a digital reminder.

\subsection{Front-End Design \& Development}
We scripted the front-end interface in HTML, CSS, \& JavaScript with background styling using a gray to black linear gradient with white features (Figure 1). The page layout featured left and right sections, where the left contained user instructions and the right contained an image upload form. A semi-transparent spinner was programmed to show during background tasks. 

The front-end JavaScript functionality captured and handled user interactions and communicated with the remote server. It also sent the image file to the server and handled the server response. Finally, it dynamically generated a download link when the server provided the expected ics file.

\begin{figure}[htbp]
    \centering
    \includegraphics[width=0.5\textwidth]{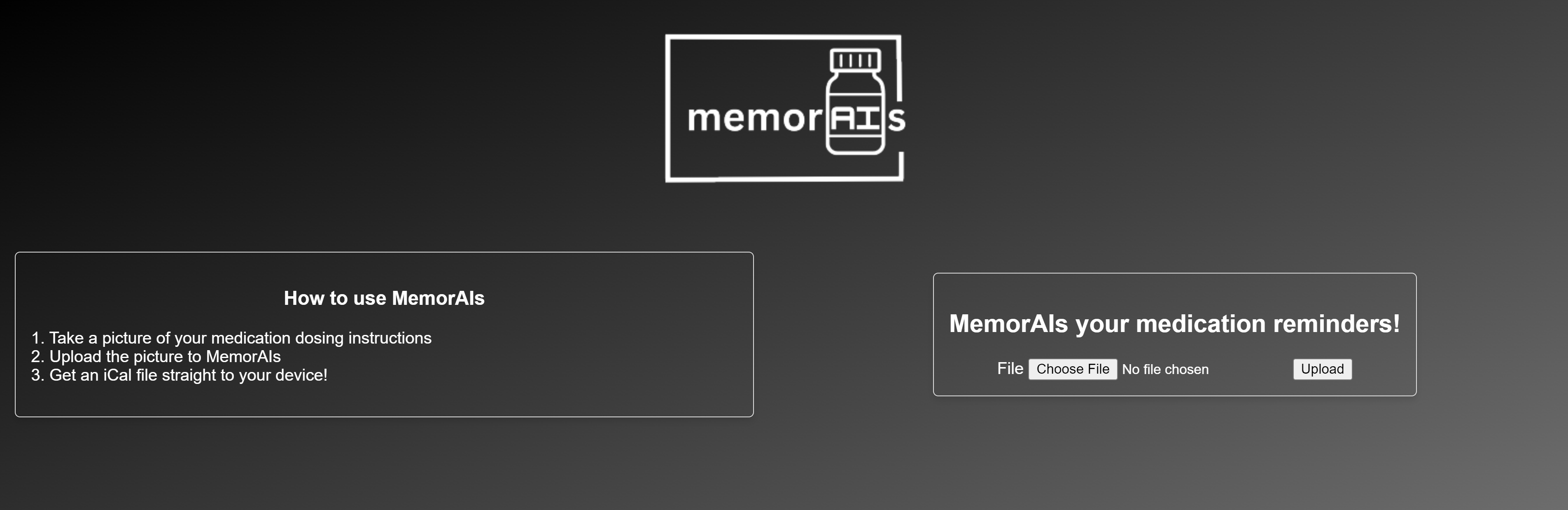}
    \caption{memorAIs User Interface.}
    \label{fig}
\end{figure}

\subsection{Back-End Development \& Rule-Based Text Processing}
We developed the back-end using Python and employed a Flask server for the memorAIs application. The back-end architecture consisted of a well-organized set of scripts, configuration files, and modules, tailored to facilitate different facets of our application, such as supporting the Flask server, implementing web services, managing the application programming interface (API), and handling data.

The rule-based text processing was handled through a single script that extracted text from images of medication instructions from medication containers and generated medication reminder events in ics format. This process was broken up into four parts: 1.) text extraction, 2.) text preprocessing, 3.) text interpretation, and 4.) information formatting and return.

\subsection{Text Extraction}
Text extraction from images uploaded on the platform were handled via optical character recognition (OCR). OCR is a computer vision task that converts text on images or printed documents into machine-readable text \cite{b19}. There are several ways to implement OCR, including rule-based OCR, template matching, feature-based OCR, and machine learning OCR to name a few. While developing memorAIs, we tested two OCR implementation packages from the Python library: 1.) pytesseract, a Python wrapper for Google's Tesseract-OCR Engine which uses a blend of traditional computer vision techniques and modern deep learning methods for character recognition \cite{b20}, and 2.) PaddleOCR, an OCR implementation which leverages PaddlePaddle’s deep learning framework \cite{b21} (Figure 2).

We chose to use the PaddleOCR package for memorAIs, citing higher accuracy and the package’s ability to return String bounding box coordinates for each detected text element. We determined that this feature would be helpful in our later implementation of regular expressions that would consider character coordinates in placement probabilities.

\begin{figure}[htbp]
    \centering
    \includegraphics[width=0.5\textwidth]{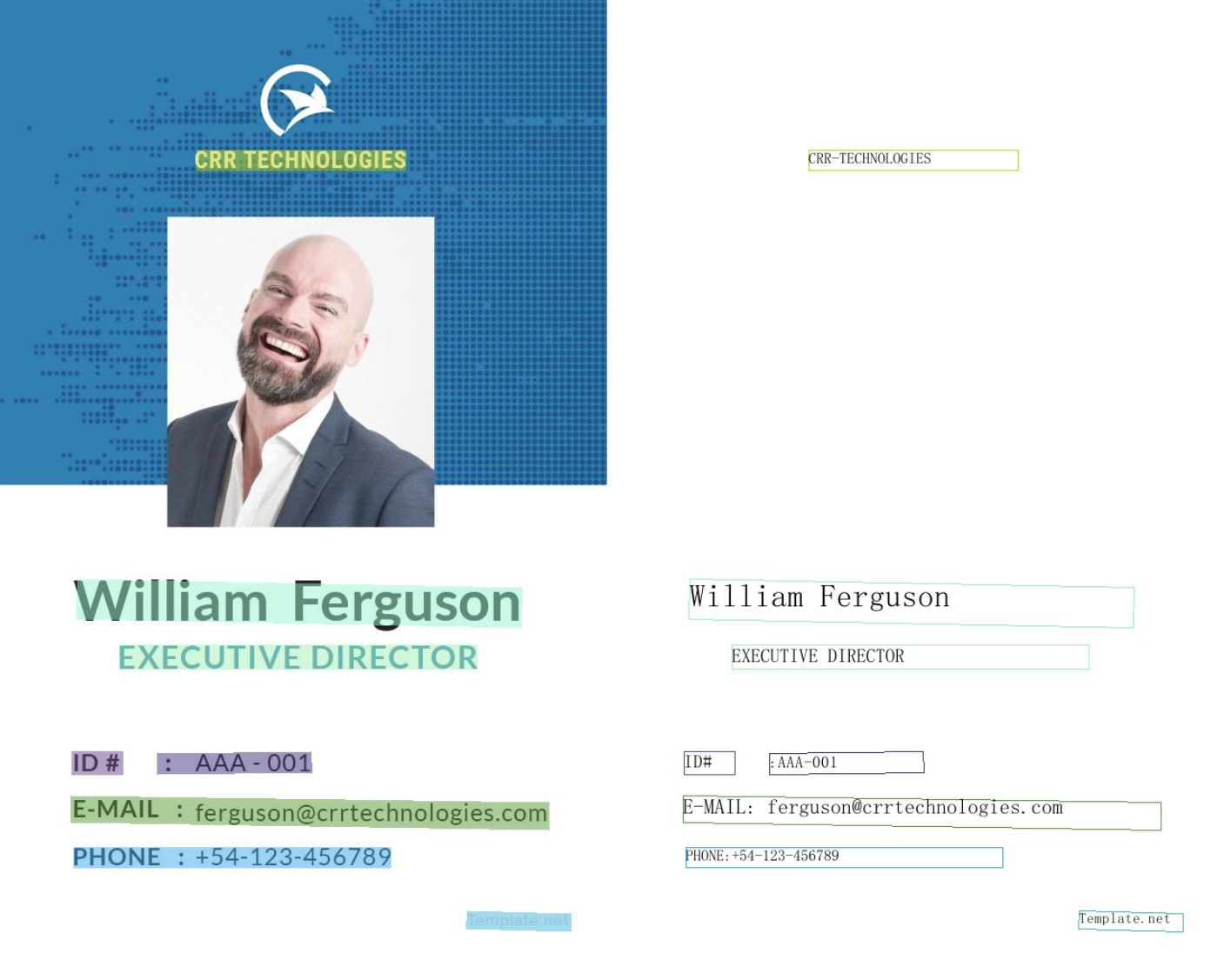}
    \caption{PaddleOCR Bounding Box Example (from PaddleOCR GitHub)}
    \label{fig}
\end{figure}

\subsection{Text Preprocessing}
Text was preprocessed in two ways: 1.) bounding box coordinate comparison, and 2.) standardized character representation. Leveraging the bounding box coordinate information provided by the PaddleOCR package, we wrote a function to compare the coordinates of all detected bounding boxes to determine the order in which they appear on the bottle. Next, we converted all String representations of numbers to numeric form (e.g., "two" as 2) for uniform representation.

\subsection{Text Interpretation}
In recognition of the fact that U.S. prescription pill bottle instructions tend to contain similar text elements concerning medication intake time, duration, and frequency, we opted for a regular-expression-based approach to text interpretation. To inform medication reminder components, we developed 20 regular expressions as frequency indicators and nine regular expressions to determine the duration of intake.

Example frequency regular expressions included:
\begin{itemize}
\item \texttt{"regex": r"every other day", "frequency": 0.5, "frequency\_unit": "days"}
\item \texttt{"regex": r"twice per day", "frequency": 2, "frequency\_unit": "days"}
\item \texttt{"regex": r"every night", "frequency": 1, "frequency\_unit": "days", "time\_of\_days": ["evening"]}
\end{itemize}

Example frequency duration expressions included:
\begin{itemize}
\item \texttt{"regex": r"for ([0-9]+) days", "duration\_unit": "days"}
\item \texttt{"regex": r"after ([0-9]+) months", "duration\_unit": "months"}\\
\end{itemize}

The general rule-based classification algorithm was as follows:\\

\textbf{Input:} Preprocessed Text (\texttt{label\_text})\\
\textbf{Output:} Medication Schedule Parameters\\

\textbf{Parameters:}
\begin{itemize}
    \item[\textbullet] \textbf{frequency:} How often the medication should be taken
    \item[\textbullet] \textbf{frequency unit:} Unit of frequency (e.g., hours, days, weeks)
    \item[\textbullet] \textbf{duration:} Duration for which the medication should be taken
    \item[\textbullet] \textbf{duration unit:} Unit of duration (e.g., days, weeks, months)
    \item[\textbullet] \textbf{time of day:} Times of day for medication intake
\end{itemize}

\textbf{Rule-based Classification:}
\begin{itemize}
    \item[\textbullet] For each rule in \texttt{frequency\_indicator}, \texttt{duration\_indicator}:
    \begin{itemize}
        \item[\textendash] Apply the regular expression to \texttt{label\_text}
        \item[\textendash] If a match is found:
        \begin{itemize}
            \item[\textasteriskcentered] Update the corresponding parameter based on the rule
        \end{itemize}
    \end{itemize}
\end{itemize}

\subsection{Source Code \& Materials}
\noindent
Front-End GitHub Repository: \href{https://tinyurl.com/f3enrp4d}{Click here}\\
Back-End GitHub Repository: \href{https://github.com/utkarshsingh99/Memorais}{Click here}\\
Web Application: \href{https://memorais.netlify.app/}{Click here}\\

\textbf{Note:} In order to be functional, the current memorAIs implementation relies on making a call to a server that is currently turned off. If the product gains traction, we will search for a long term server solution so that anyone, anywhere can use it.

\subsection{Information Formatting \& Return}
Once the text had been interpreted based on our indicator rules, the information was stored as values in a JSON file. From there, an empty Calendar object was created and values extracted from the JSON file were used to map values to Calendar object format using the icalendar Python library. The populated calendar object, configured with the appropriate reminder information (see examples in Figure 3) was returned and sent to the front-end, which then generated the ics download link to the user’s device.

\begin{figure}[htbp]
    \centering
    \includegraphics[width=0.5\textwidth]{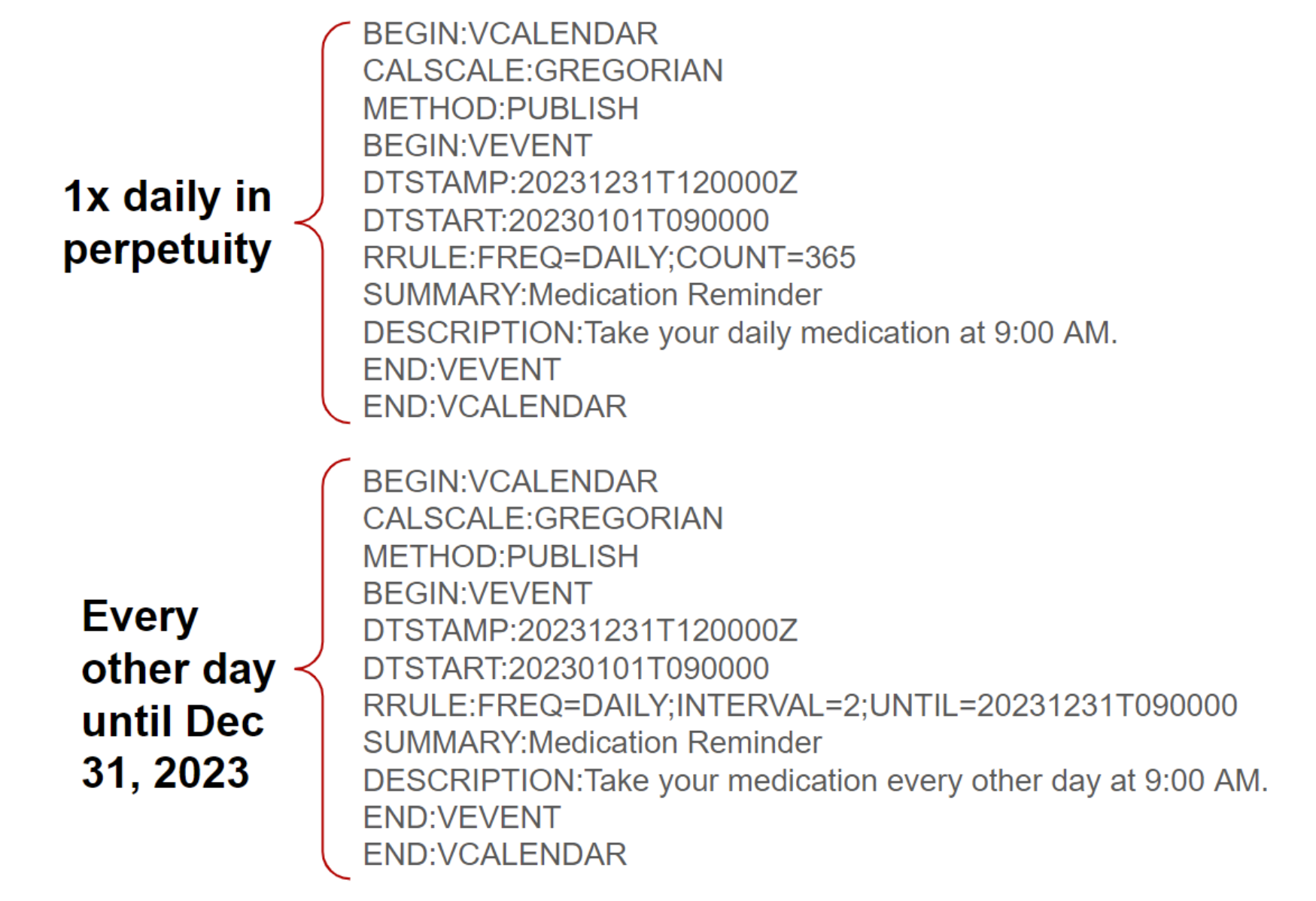}
    \caption{Example ics File Content}
    \label{fig}
\end{figure}

\section{Discussion}
\subsection{Limitations}
There is no shortage of limitations in short-term hackathon projects; however, one of the most prominent included our strategy for text processing. Our text processing methodology was conducted via hardcoded regular expressions as opposed to more generalizable approaches, such as natural language processing methods. Because of this, our approach lacked generalizability and was only useful up to a certain point of instruction phrasing complexity. Additionally, using a language model would have been useful for more meaningfully integrating other indicators, aside from frequency and duration, including medication name, and intake routine details.

Another substantial limitation was that our approach lacked a way to remove or overwrite old ics files when they became outdated. One of the express appeals of a solution like memorAIs is the ability to reduce burden when medications are updated. In a more developed implementation, we would have included a way to remove outdated files.

\subsection{Conclusion}
Memory-based medication non-adherence is an unsolved problem that is responsible for considerable disease burden in the United States. Digital medication intake reminder solutions with minimal onboarding requirements that are usable at the point of medication acquisition may help to alleviate this problem by offering a low barrier way to help people remember to intake their medications. In this paper, we proposed memorAIs, a digital medication intake reminder solution that mitigated onboarding friction by leveraging OCR for text extraction from medication bottles and rule based expressions for text processing to create configured medication reminders as local device calendar invitations. While there were several limitations to our implementation, including the lack of generalizability of our text interpretation beyond hardcoded regular expressions and an inability to replace outdated ics files, memorAIs was awarded the first-place Patient Safety award at the 2023 DivHacks Hackathon by the Patient Safety Technology Challenge, sponsored by the Pittsburgh Regional Health Initiative. In a formal implementation of memorAIs as a viable product, we would improve our text processing and interpretation methodology, integrate a way to remove outdated ics files from the user’s device, and improve our UI for a more seamless and enjoyable user experience.

\section*{Acknowledgment}
We would like to thank the 2023 DivHacks team at Columbia University for organizing the hackathon at which memorAIs was conceived and the Pittsburgh Regional Health Initiative for awarding us with the Patient Safety award.

\end{document}